\documentclass[12pt,a4paper]{article}

\usepackage{amsmath}
\usepackage{amsfonts}
\usepackage{amssymb}
\usepackage{amsthm}
\usepackage{slashed} 
\usepackage{graphicx}
\usepackage{epstopdf}
\usepackage[dvipsnames,svgnames,x11names]{xcolor}
\usepackage{hyperref}
\usepackage[figure,table]{hypcap} 

\newcommand{\OO}{\mathrm{O}}      
\newcommand{\SO}{\mathrm{SO}}

\newcommand{\dd}{\mathrm{d}}

\newcommand{\R}{\mathbb{R}}   
\newcommand{\C}{\mathbb{C}}   
\newcommand{\Z}{\mathbb{Z}}

\DeclareMathOperator{\Realpart}{Re}

\numberwithin{equation}{section}

\begin{document}

\title{A topological state sum model for a scalar field on the circle}

\author{Steven Kerr \\
stevenkerr2@gmail.com
}

\maketitle

\begin{abstract} 
This paper is a follow-up to a previous paper on fermions \cite{fermions}. A simple state sum model for a scalar field on a triangulated $1$-manifold is constructed. The model is independent of the triangulation and gives exactly the same partition function as the continuum functional integral with zeta function regularisation. For a certain choice of gauge group, the state sum model on the circle is equivalent to the path integral for the simple harmonic oscillator.
\end{abstract}

\section{Introduction}

Lattice discretisation procedures are often used to define the partition functions of quantum theories. The resulting models are called state sum models, because the functional integral become a sum over a discrete set of states. These models typically depend on the particular lattice structure that is chosen, but in the limit as the lattice becomes increasingly fine, the model becomes a better approximation to the continuum theory.

In topological quantum field theories, the model is independent of the choice of lattice structure, which is typically taken to be a triangulation of the spacetime manifold. Such theories are of interest in quantum gravity because the partition function has the same symmetry as the classical theory, namely spacetime diffeomorphism symmetry.
Realistic quantum gravity models should include matter, and so it is important to consider models which incorporate matter while maintaining triangulation independence. With the recent discovery of the Higgs boson at the LHC \cite{Higgs}, it is now clear that scalar fields are an important part of our description of fundamental physics. This is our motivation for studying state sum models which include scalar fields.

In this paper, a simple one dimensional model of a scalar field on a $1$-manifold is constructed. In section \ref{sec: state sum}, the definition of the state sum model for a real scalar field on the interval and circle is developed. The model is independent of the triangulation and depends only on the holonomy of the gauge field. However, introduction of a mass term in a `na\"{i}ve' way breaks the triangulation independence. In section \ref{sec: interpretation}, it is shown that the state sum model has a discrete action that has the continuum action for a minimally coupled scalar field as its continuum limit. In section \ref{sec: functional}, the functional integral for the continuum theory on the circle is evaluated using zeta function regularisation, and it is shown to be exactly equal to the result from the state sum model for the gauge group $\OO(n)$. If the gauge group is chosen to be the group of strictly positive numbers under multiplication, the state sum model on the circle is equivalent to the path integral for the harmonic oscillator, or equivalently the massive scalar field in one dimension. Thus it is possible to introduce a mass term into the model while maintaining triangulation independence if the mass parameter is treated in the same fashion as a gauge field. In the standard treatment of the path integral for the harmonic oscillator, the partition function is calculated as the limit of a discrete model that is not triangulation independent. The state sum model presented here has the virtue that it is triangulation independent and exactly equal to the partition function of the harmonic oscillator once the appropriate gauge group has been chosen.

There is not much work on state sum models which incorporate scalar fields in the literature, perhaps because it was not known until recently if scalar fields exist in nature. However, the quantum mechanics of a particle in $n$ dimensions is in fact equivalent to a one dimensional quantum field theory of an $n$-plet of scalar fields, in precisely the same way that the quantum mechanics of a string in $n$ dimensions is equivalent to a two dimensional quantum field theory of an $n$-plet of scalar fields on the worldsheet. The path integral quantisation of the harmonic oscillator has been well explored; however, in this paper we develop an alternate approach which is more in the spirit of quantum field theory. This work complements our previous work on fermions \cite{fermions}.

It is an interesting question as to what extent the model here can be generalised to higher dimensions. We leave this for future investigation.

\section{The state sum model}\label{sec: state sum}

\begin{figure}[h!]  
\begin{center}
\includegraphics[scale=0.7]{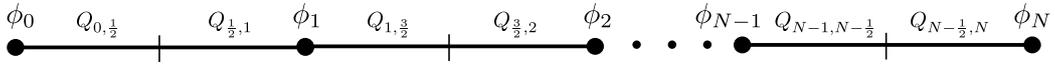}
\end{center} 
\caption{The state sum model for a scalar field on the interval.}\label{interval figure}
\end{figure}

Start with an oriented interval $[0,l]$ of length $l$, triangulated with $N+1$ vertices. The vertices are decorated with variables $\phi_i, i=0\ldots N$, each of which is a vector in $\R^n$. The edge connecting the $i$-th and $(i+1)$-th vertices is further subdivided into two segments by a vertex at its centre labelled by $i+\frac{1}{2}$. Each segment with initial vertex $\alpha$ and final vertex $\beta$ is decorated with a real $n \times n$ matrix $Q_{\alpha,\beta}$. Indeed we will use the more general notation that $Q_{\alpha,\beta}$ is equal to the product of the matrices connecting vertices $\alpha$ and $\beta$, which need not be adjacent, in the order determined by the orientation. These matrices satisfy $Q_{\alpha,\beta} = Q_{\beta,\alpha}^{-1}$. The length of each edge is $\Delta t = \frac{l}{N}$. For now we assume that the matrices $Q_{\alpha,\beta}$ are orthogonal. The situation is depicted in figure \ref{interval figure}.

For $N=1$, i.e. a single edge, the state sum model is

\begin{align}
\Z_1 = \left( \frac{1}{2\pi \Delta t} \right)^{\frac{n}{2}} e^{ -\frac{1}{2\Delta t} (Q_{\frac{1}{2},0} \phi_0 - Q_{\frac{1}{2},1} \phi_1  )^2},
\end{align}
with $\Delta t = l$. Gluing two edges together is carried out by multiplying their respective partition functions together and integrating over the variable associated to the interior vertex,

\begin{align}
\Z_2 = \left( \frac{1}{2\pi \Delta t} \right)^{n} \int \dd \phi_1 \; e^{ -\frac{1}{2\Delta t} (Q_{\frac{1}{2},0} \phi_0 -  Q_{\frac{1}{2},1} \phi_1 )^2} e^{ -\frac{1}{2\Delta t} (Q_{\frac{3}{2},1}\phi_1 -  Q_{\frac{3}{2},2} \phi_2)^2}, \label{z2}
\end{align} 
where the integral is the Lebesgue integral over $\R^n$, and now $N=2$, $\Delta t = \frac{l}{2}$. This can be evaluated using the following lemma,

\begin{align}
 \int \dd u \; e^{ -\frac{1}{2a} (x  - M_1 u)^2 } e^{ -\frac{1}{2b} (M_2 u  -  y)^2} = \left(\frac{2\pi ab}{a+b}\right)^{\frac{n}{2}} e^{ -\frac{1}{2(a+b)} (M_1^T x -  M_2^T y)^2},
\end{align}
where $x,y,u \in \R^n$, $a,b>0$ are real numbers, $M_1, M_2$ are real, orthogonal $n \times n$ matrices, and the superscript $T$ denotes the matrix transpose. This can be proved by expanding the brackets and using gaussian integration.

Applying this to \eqref{z2} results in

\begin{align}
\Z_2 &= \left( \frac{1}{4 \pi \Delta t} \right)^{\frac{n}{2}} e^{ -\frac{1}{4\Delta t} (Q_{1,0}\phi_0  - Q_{1,2} \phi_2)^2}. \label{Pachner}
\end{align}
Gluing $N$ edges together in this way yields the definition of the state sum model on an interval triangulated with $N+1$ vertices,

\begin{align}
\Z_{[0,l]} &= \left( \frac{1}{2\pi \Delta t} \right)^{\frac{Nn}{2}} \int \prod_{i=1}^{N-1} \dd \phi_i \; e^{ -\frac{1}{2\Delta t} \sum_{i=0}^{N-1} (Q_{i+\frac{1}{2},i}\phi_{i} - Q_{i+\frac{1}{2},i+1} \phi_{i+1})^2} \label{interval0} \\
&= \left( \frac{1}{2\pi N \Delta t} \right)^{\frac{n}{2}} e^{ -\frac{1}{2 N \Delta t} (Q_{m,0} \phi_0  -  Q_{m,N} \phi_{N})^2}  \\
&=  \left( \frac{1}{2\pi l} \right)^{\frac{n}{2}} e^{ -\frac{1}{2l} ( Q_{m,0}\phi_0  -  Q_{m,N} \phi_{N})^2}. \label{scalar interval}
\end{align}
Here $m$ denotes the vertex at the midpoint of the interval.

The state sum model on the circle is obtained by identifying $\phi_{N}=\phi_0$ in \eqref{interval0} and integrating,

\begin{align}
\Z_{S^1} &= \left( \frac{1}{2\pi \Delta t} \right)^{\frac{Nn}{2}} \int \prod_{i=0}^{N-1} \dd \phi_i \; e^{ -\frac{1}{2\Delta t} \sum_{i=0}^{N-1} (Q_{i+\frac{1}{2},i}\phi_{i} - Q_{i+\frac{1}{2},i+1} \phi_{i+1})^2}. \\
&= \left( \frac{1}{2\pi \Delta t} \right)^{\frac{Nn}{2}} \int \dd \Phi e^{-\frac{1}{2\Delta t} (M\Phi)^2}, 
\end{align}
where $\Phi = \oplus_{i=0}^{N-1} \phi_i$, and

\begin{align}
M = \left( \begin{array}{ccccc}
Q_{\frac{1}{2},0} & -Q_{\frac{1}{2},1} & & & \\
& Q_{\frac{3}{2},1} & -Q_{\frac{3}{2},2} & & \\
& & \ddots & \ddots & \\
& & & Q_{N-\frac{3}{2},N-2} & -Q_{N-\frac{3}{2},N-1} \\
-Q_{N-\frac{1}{2},N} & & & & Q_{N-\frac{1}{2},N-1} \\
\end{array} \right)
\end{align}
On the circle, the orthogonality assumption for the $Q$'s can be weakened; they can now be assumed to be invertible only. The integral can be evaluated with a change of variables, yielding

\begin{align}
\Z_{S^1} = \frac{1}{|\det M|}.
\end{align}
The determinant can be computed iteratively using the following fact,

\begin{align}
\det\left( \begin{array}{cc}
A & B\\
C & D \\
\end{array} \right) = \det \left( \begin{array}{cc}
1 & B\\
0 & D \\
\end{array} \right) \left( \begin{array}{cc}
A-BD^{-1}C & 0 \\
D^{-1}C & 1 \\
\end{array} \right) = \det D \det(A - BD^{-1}C),
\end{align}
where $A$, $B$, $C$, $D$ are $p\times p$, $p \times q$, $q \times p$ and $q \times q$ matrices, respectively, and $D$ is invertible. The result is

\begin{align}
\Z_{S^1} = \left| \left( \prod_{i=0}^{N-1} \det Q_{i+\frac{1}{2}, i} \right) \det(1-Q) \right|^{-1}, \label{scalar circle}
\end{align}
where $Q = \prod_{i=0}^{N-1} Q_{i,i+1}$ is the holonomy around the circle.

The partition function \eqref{scalar interval} on the interval has the property of triangulation independence. That is, it is independent of $N$, the number of sides of the polygon. This is also true of the partition function \eqref{scalar circle} on the circle, provided some further stipulations are placed on the $Q$'s.

In the case where the $Q$'s are elements of $\SO(2)$, and $Q=e^{\theta X}$ with $X=\left( \begin{array}{cc}
0 & -1\\
1 & 0 \\
\end{array} \right)$, $\theta \neq 0$, the partition function \eqref{scalar circle} is given by

\begin{align}
\Z_{S^1} = \frac{1}{4 \sin^2  \frac{\theta}{2}} \label{so(2) result}.
\end{align}
More generally, if $Q$ is an orthogonal matrix, it may be diagonalised to the following canonical form

\begin{align}
\left( \begin{array}{cccccc}
 R_1 & & & & & \\
 & \ddots & & & & \\
 & & R_k & & &\\
 & & & \pm 1 & &  \\
 & & & & \ddots & \\
 & & & & & \pm 1 \end{array} \right), 
\end{align}
where the $R_i$, $i=1..k$ are independent $2 \times 2$ rotation blocks. It is clear that under certain circumstances the denominator in \eqref{scalar circle} can be zero, in which case the model is not defined. This occurs if $Q$ is odd-dimensional. Thus this scenario is excluded from consideration. Then the partition function \eqref{scalar circle} decomposes as the product of a number of $\SO(2)$ theories.

In the case where $Q_{i,i+\frac{1}{2}} = Q_{i+\frac{1}{2},i+1} = e^{\frac{\psi}{2N}} I \; \forall i $ with $\psi \in \R \setminus \{0\}$ and $I$ the $n \times n$ identity matrix, the partition function \eqref{scalar circle} is

\begin{align}
\Z_{S^1} = \left( \frac{1}{\left|2 \sinh \frac{\psi}{2}\right|} \right)^n \label{SHO circle result}.
\end{align}

The state sum model presented here can be straightforwardly generalised to the case where the $\phi_i$ are complex vectors. In this case, where it occurs the orthogonality requirement for the matrices $Q_{\alpha,\beta}$ is replaced by unitarity, and the partition functions \eqref{scalar interval}, \eqref{scalar circle} are the same but for twice as many real degrees of freedom.

\section{Interpretation of the state sum model}\label{sec: interpretation}

The state sum model of the previous section may be interpreted as the partition function of a minimally coupled, real scalar field theory.

The $\phi_i$'s are interpreted as a discrete sampling of a continuous real scalar field $\phi(t)$ on the circle, with $t\in [0,l)$, so that

\begin{align}
\phi_j = \phi\left(j\Delta t \right).
\end{align}
The interval connecting vertices $\alpha$ and $\beta$ is assigned the matrix $Q_{\alpha,\beta}$, which is interpreted as the parallel transporter for the $\phi$ field along that interval. Then $Q_{\alpha,\beta} = e^{\int_{\alpha \Delta t}^{\beta \Delta t} \dd t  A(t)} $, with $A(t)$ the gauge field on the circle. Up to a minus sign, the argument of the exponent in \eqref{interval0} may be seen to be a lattice discretisation of the usual action for a scalar field,

\begin{align}
\hat{S} &=  \frac{\Delta t }{2} \sum_{i=0}^{N-1} \left(\frac{Q_{i+ \frac{1}{2},i }\phi_{i} - Q_{i+\frac{1}{2},i+1} \phi_{i+1} }{\Delta t}\right)^2 .\label{discrete action}
\end{align}
The limit $N \rightarrow \infty$, or equivalently $\Delta t \rightarrow 0$, while keeping $l$ constant can be evaluated,

\begin{align}
\lim_{\Delta t \rightarrow 0} \hat{S} &= \lim_{\Delta t \rightarrow 0} \frac{\Delta t }{2} \sum_{i=0}^{N-1} \left(\frac{Q_{i+ \frac{1}{2},i }\phi_{i} - Q_{i+\frac{1}{2},i+1} \phi_{i+1} }{{\Delta t}}\right)^2  \nonumber  \\
&= \frac{1}{2} \int_0^l \dd t \; (D\phi)^2, \label{continuum limit}
\end{align}
where $D = \frac{\dd}{\dd t}  + A$ is the covariant derivative. This is just the usual continuum action for a scalar field. Thus the partition function \eqref{interval0} may be interpreted as that of a real, minimally coupled scalar field.

It is possible to add a mass term to the action \eqref{discrete action},

\begin{align}
\hat{S}_m = \frac{1}{2} m^2 \Delta t \sum_{i=0}^{N-1} \phi_i^2 . \label{naive mass term}
\end{align}
However, the resulting partition functions on the interval and circle are no longer triangulation independent.

If the gauge group is taken to be isomorphic to the abelian group of strictly positive real numbers under multiplication so that $A(t) \sim I$, then

\begin{align}
\frac{1}{2} \int_0^l \dd t \; (D\phi)^2 = \frac{1}{2} \int_0^l \dd t \left[ \left( \frac{\dd \phi}{\dd t} \right)^2 + \alpha^2 \phi^2 \right]. \label{massive scalar}
\end{align}
Here the gauge freedom has been used to transform the gauge field $A(t)$ so that it is everywhere equal to a constant $\alpha I$. Identifying $\phi(t)$ with the $n$-dimensional position vector $x(t)$ and $\alpha$ with the spring constant reveals that this is precisely the Euclidean action for the simple harmonic oscillator. Alternatively the action may be viewed as that of a massive scalar field upon identifying $\alpha$ with the mass parameter. Thus it is possible to introduce a mass term into the state sum model if the mass parameter is treated as an element of the Lie algebra $\R$. The corresponding matrices $Q_{\alpha,\beta}$ furnish an $n$-dimensional representation of the abelian group of strictly positive numbers under multiplication, as is the case in \eqref{SHO circle result}. In this way it is possible to include a mass term in the state sum model on the circle while maintaining triangulation independence.

\section{Comparison to functional integral}\label{sec: functional}

In this section, we compute the zeta function regularised partition function of the continuum theory for the gauge group $\SO(2)$, and show that it is equal to the result from the state sum model. 

Zeta function regularisation is a method of regularising divergent products. It can be used to define the determinant of operators on infinite dimensional spaces, i.e. differential operators. The motivating observation is the following: Let $L$ be a Hermitian, strictly positive operator in a finite-dimensional Hilbert space with no eigenvalues that are zero.  The zeta function $\zeta_{L} (s)$ of $L$ is defined for $s\in\C$ by

\begin{align}
\zeta_{L} (s) = \sum_k \frac{1}{\lambda_k^s},
\label{eq:zetaLdef}
\end{align}
where $\lambda_k$ are the eigenvalues of~$L$. As $L$ has a finite number of positive eigenvalues,  $\zeta_{L} (s)$ is well defined and holomorphic in~$s$. An elementary computation shows that 

\begin{align}
\det L = \prod_k \lambda_k = e^{- \zeta_L'(0) } . 
\label{eq:detLdef}
\end{align}

The point of \eqref{eq:detLdef} is that the expression on the right hand side may, under certain circumstances, be taken as a definition of $\det L$ even when the Hilbert space is infinite dimensional. We require the spectrum of $L$ to be discrete, and the sum in \eqref{eq:zetaLdef} must converge for sufficiently large $\Realpart s$ to define $\zeta_{L} (s)$ as a function that can be analytically continued to $s=0$.  The analytic continuation in $s$ provides a prescription for regularising the divergent product~$\prod_k \lambda_k$.

This is useful in quantum field theory because one can use zeta function regularisation to define the partition function of certain theories. So for a minimally coupled scalar field theory, we may define

\begin{align}
\Z &= \int \mathcal{D} \phi \; e^{- \frac{1}{2} \int \phi L \phi }\\
&:= \frac{1}{{\sqrt{\det L}}}, \label{continuum partition function}
\end{align}
where the square root is the positive square root. The differential operator $L$ gives the classical action in the exponent. In the case of a real scalar field minimally coupled to an $\OO(n)$ gauge field on the circle, we have

\begin{align}
S &= \frac{1}{2} \int_0^l \dd t \; (D\phi)^2 \\
&=  \frac{1}{2} \int_0^l \dd t \; \phi \left( -\frac{\dd^2}{\dd t^2} -2A \frac{\dd}{\dd t} - A^2 \right) \phi, \label{action1}
\end{align}
which gives $L= -\frac{\dd^2}{\dd t^2} -2A \frac{\dd}{\dd t} - A^2$. In order to compute $\det L$, we first need to know the eigenvalues of $L$. We will take the gauge group to be $\SO(2)$, so that $A = \alpha \left( \begin{array}{cc}
 0 & -1 \\
 1 & 0 \\
 \end{array} \right)$. Using the gauge freedom, we may take $\alpha$ to be a constant which we denote by $\frac{2\pi a}{l}$, and by a further gauge transformation we may take $a \in [0,1)$. Then the eigenvalue equation is

\begin{align}
\left( \begin{array}{c}
 -\frac{\dd^2 v_1}{\dd t^2} +2\alpha \frac{\dd v_2}{\dd t} + \alpha^2 v_1  \\
 -\frac{\dd^2 v_2}{\dd t^2} -2\alpha \frac{\dd v_1}{\dd t} + \alpha^2 v_2 \\
 \end{array} \right)
= 
\lambda \left( \begin{array}{c}
  v_1  \\
  v_2 \\
 \end{array} \right),
\end{align}
and we impose periodic boundary conditions, $v_{1,2} (t+l) =v_{1,2} (t) $. This has linearly independent solutions $\left( \begin{array}{c}
  \cos \frac{2\pi kt}{l}  \\
  \sin \frac{2\pi kt}{l} \\
 \end{array} \right)$ and $\left( \begin{array}{c}
  \sin \frac{2\pi kt}{l}  \\
  \cos \frac{2\pi kt}{l} \\
 \end{array} \right)$, $k \in \Z$. The eigenvalues are given by
 
 \begin{align}
 \lambda_{k_{\pm}} = \left(\frac{2\pi}{l}(k \pm a)\right)^2, \;\; k \in \Z.
 \end{align}
The determinant in \eqref{continuum partition function} is defined by 

\begin{align}
\det L = e^{- \zeta_L'(0) },
\end{align}
with 

\begin{align}
\zeta_L(s) &=  \left(\frac{2\pi}{l}\right)^{-2s} \sum_{k \in \Z} \left(  \frac{1}{(k+a)^{2s}} + \frac{1}{(k-a)^{2s}} \right) \\
&= \left(\frac{2\pi}{l}\right)^{-2s} \sum_{k \in \Z} \frac{2}{(k+a)^{2s}}. \label{zeta}
\end{align}
This may be re-written in terms of the Hurwitz zeta function,

\begin{align}
\zeta_L(s) =  \left(\frac{2\pi}{l}\right)^{-2s} \Big( 2\zeta_H(2s, a) + 2\zeta_H(2s, 1-a)\Big),
\end{align}
where $\zeta_H$ is defined by

\begin{align}
\zeta_H(s, q) = \sum_{k=0}^{\infty} \frac{1}{(k+q)^s}.
\end{align}
The Hurwitz zeta function can be analytically continued to remove the pole at $s=0$. Then using 25.11.13, 25.11.18 
and 5.5.3 in~\cite{dlmf}, we have

\begin{align}
\det L = (4 \sin^2 \pi a)^2,
\end{align}
and the partition function \eqref{continuum partition function} is

\begin{align}
\Z = \frac{1}{4 \sin^2 \pi a}.
\end{align}
The holonomy is given by $Q=e^{2\pi a}$, and identifying $\theta = 2\pi a$ gives precisely the result \eqref{so(2) result} from the state sum model. Thus the zeta function regularisation and the lattice regularisation are seen to be equivalent.

The action for the Euclidean harmonic oscillator in $n$ spatial dimensions is

\begin{align}
S = \frac{1}{2} \int_0^l \dd t \; x(t) \left( -\frac{\dd^2}{\dd t^2} + \omega^2 \right) x(t). \label{SHO}
\end{align}
The corresponding partition function on the circle is calculated in \cite{HO},

\begin{align}
{\mathbb Z}_{S^1} = \left( \frac{1}{2\sinh\frac{\omega l}{2}}\right)^n,
\end{align}
where $\omega > 0$ is understood to be the positive square root of $\omega^2$. This result is identical to the state sum model \eqref{SHO circle result} after identifying $|\psi|=\omega l$.

\section{Discussion}
In this paper, a one dimensional state sum model for a real scalar field on the circle and interval has been constructed. The resulting partition functions are simple functions of the holonomy that are triangulation independent. I have carried out an exact calculation of the partition function in the continuum using zeta function methods, and have obtained results identical to the state sum model for the gauge group $\OO(n)$. If the gauge group is chosen to be the group of strictly positive real numbers under multiplication, the state sum model on the circle is exactly equal to the path integral for the harmonic oscillator. Normally the path integral for the harmonic oscillator is constructed as the limit of a state sum that is not triangulation independent. The model here has the added virtue of triangulation independence, and thus in many ways represents a significant simplification.

Introduction of a mass term in the most obvious way breaks the triangulation independence of the state sum model. This is in agreement with work by Rovelli \cite{Rovelli}. However, a mass term can still be accommodated while maintaining triangulation independence if the mass parameter is treated in the same manner as a gauge field.
 
The results here can be extended to the case of a complex scalar field in a straightforward way. The Lorentzian path integrals can be obtained by Wick rotation.

The most interesting generalisation of this work would be to higher dimensions. We hope that the model presented here may clarify some of the issues that arise in that case.

\section{Acknowledgements}
I thank John Barrett, Jorma Louko and Sara Tavares.


\begin{thebibliography}{99}

\bibitem{fermions}
J.~W. Barrett, S.~Kerr and J.~Louko
``A topological state sum model for fermions on the circle"
(2012)
\href{http://arxiv.org/abs/1211.4557}{arXiv:1211.4557 [math-ph]}.

\bibitem{Higgs}
ATLAS Collaboration (Georges Aad (Freiburg U.) et al.)
``Observation of a new particle in the search for the Standard Model Higgs boson with the ATLAS detector at the LHC ",
Phys.Lett. B716 (2012) 1-29 
\href{http://arxiv.org/abs/1207.7214}{arXiv:1207.7214 [hep-ex]}. 

\bibitem{Rovelli}
Carlo Rovelli
``Discretizing parametrized systems: the magic of Ditt-invariance"
(2011)
\href{http://arxiv.org/abs/1107.2310}{arXiv:1107.2310 [hep-lat]}. 

\bibitem{dlmf}
{\it Digital Library of Mathematical Functions\/} 
(National Institute of Standards and Technology, 2011-08-29), 
{\tt \url{http://dlmf.nist.gov/}.} 

\bibitem{HO}
Vadim Kaplunovsky
``Path integral for the harmonic oscillator" lecture notes
{\tt \url{http://bolvan.ph.utexas.edu/~vadim/classes/2004f.homeworks/osc.pdf}.} 

\end{thebibliography}
\end{document}